\documentclass[aip,jmp,amsmath,amssymb,preprint]{revtex4-1}

\usepackage{graphicx}
\usepackage{dcolumn}
\usepackage{bm}
\usepackage{xcolor}
\usepackage{amsfonts,amsmath,amsthm,amssymb,mathtools}
\usepackage{longtable}

\begin{document}
\preprint{AIP/JMP}

\title{Calculation of STOs electron repulsion integrals by ellipsoidal expansion and large-order
approximations}

\author{Micha\l~Lesiuk}
\email{lesiuk@tiger.chem.uw.edu.pl.}
\affiliation{Faculty of Chemistry, University of Warsaw, Pasteura 1, 02-093 Warsaw, Poland}

\date{\today}

\begin{abstract}
For general two-electron two-centre integrals over Slater-type orbitals (STOs), the use of the Neumann expansion for the
Coulomb interaction potential yields infinite series in terms of few basic functions. In many important cases the
number of terms necessary to achieve convergence by a straightforward summation is large and one is forced to calculate
the basic integrals of high order. We present a systematic approach to calculation of the higher-order terms in the
Neumann series by large-order expansions of the basic integrals. The final expressions are shown to be transparent and
straightforward to implement, and all auxiliary quantities can be calculated analytically. Moreover, numerical
stability and computational efficiency are also discussed. Results of the present work can be used to speed up
calculations of the STOs integral files, but also to study convergence of the Neumann expansion and develop appropriate
convergence accelerators.
\end{abstract}

\maketitle

\section{Introduction}
\label{sec:intro}

From a purely theoretical point of view, Slater-type orbitals \cite{slater30,slater32} (STOs) constitute a more
convenient basis set for calculations in molecular physics than the widely used Gaussian-type orbitals \cite{boys50}
(GTOs).
In fact, unlike GTOs, STOs are able to satisfy the Kato's cusp condition \cite{kato57} at the electron-nucleus
coalescence points and their exponential decay at large electron-nucleus distances coincides with the asymptotic form of
the electronic density \cite{agmon82} (if nonlinear parameters are suitably chosen). Only severe difficulties in
calculation of the electron repulsion integrals made the use of STOs drastically limited. Nonetheless, a considerable
interest remained in this field
\cite{jones94,barnett00,barnett002,safouhi04,safouhi06,rico00a,rico00b,rico01,smiles,hoggan09,hoggan10a,stop,
pachucki09,pachucki12a,lesiuk12,maslen90,harris02,belenruiz08,belenruiz09a,belenruiz09b,hoggan10b,belenruiz11}.

In recent three paper series \cite{lesiuk2014a,lesiuk2014b,lesiuk2015} calculation of the STOs
integrals has been reconsidered and new analytical or seminumerical methods for their computation have been proposed.
This allowed to perform calculations for the beryllium dimer with STOs basis sets up to sextuple $\zeta$ quality,
reaching the so-called spectroscopic accuracy (few wavenumbers, $\mbox{cm}^{-1}$). Additionally, it was found that the
Coulomb, $(aa|bb)$, and hybrid, $(aa|ab)$, integrals are not troublesome and are computed with a decent accuracy and
speed for a reasonable range of nonlinear parameters (and quantum numbers). Calculation of the exchange integrals,
$(ab|ab)$, is more involved. The Neumann expansion of the interaction potential, which is the method of choice, gives
rise to infinite series. In many important cases the required accuracy is obtained after summing 20-30 terms. However,
there are situations where a larger number of terms is necessary to achieve convergence which makes calculations
significantly more expensive. This is one of the major reasons for the STOs vs. GTOs gap in the computational timings.

It seems reasonable to expect that the higher-order terms in the Neumann expansion do not need to be computed with
general techniques but a suitable large-order expansion can be devised. This would allow to reduce the
computational burden significantly, as the asymptotic expansions of such kind are typically more robust than the general
expressions. Therefore, the main purpose of this paper is to derive systematic large-order approximations of all basic 
quantities appearing in the Neumann expansion of the STOs exchange integrals, and provide necessary numerical tests. 
Resulting expressions can be readily incorporated into existing STOs integral codes.

Since the present paper is concentrated solely on the Neumann expansion of the interaction potential with application to
the STOs electron repulsion integrals, a brief survey of the literature on this topic is mandatory. Relevant
mathematical details will be given in the next section. Possibly the first method utilising the Neumann expansion was
reported by Ruedenberg \cite{ruedenberg51,ruedenberg56} who introduced general expressions based on charge distributions
of both electrons. Later, this approach was extended by applying a straightforward numerical integration
\cite{ruedenberg64a,ruedenberg64b} to avoid several difficult analytic rearrangements. Kotani \cite{kotani55} provided
many tools and expressions enabling fully analytical (albeit recursive) techniques to be used. 
Recursive approach was later pursued by Harris \cite{harris60} who invoked the theory of spherical Bessel functions to
simplify the existing theory and discovered many useful additional relations. A considerable interest remained in the
field despite GTOs were clearly taking over the role of routine basis set in quantum chemistry. Many changes
were introduced in how individual terms in the Neumann expansion are computed. They were aimed at improving the
efficiency, accuracy or generality of the algorithms; the works of Yasui and Saika \cite{yasui82}, and Fern\'{a}ndez
Rico \emph{et al.} \cite{rico89,rico92,lopez94,rico94,rico94b} are notable examples. Later, Maslen and Trefry
\cite{maslen90} utilised an approach based on the hypergeometric function which enabled to derive closed-form
succinct analytical expressions for all necessary quantities. Despite those expressions were marred with numerical 
instabilities, it was a considerable progress at the time. Harris \cite{harris02} pursued the analytical approach of
Maslen and Trefry, introduced considerable simplifications and several new expressions which allow more stable
calculations of several auxiliary quantities. 

This paper is organised as follows. In Sec. \ref{sec:pre} we introduce the notation and recall relevant expressions
from the previous works. In Sec. \ref{sec:expl} we introduce the large-order asymptotic expansion for the functions
$L_\mu$ and verify the main results numerically. The corresponding expansion for the functions $W_\mu$ is given in Sec.
\ref{sec:expw}. Finally, in Sec. \ref{sec:conclusions} we conclude our paper.

\section{Preliminaries}
\label{sec:pre}

Let us consider a diatomic molecule placed on the $z$ axis symmetrically around the origin. Slater-type
orbitals (STOs) have the following generic form
\begin{align}
\label{sto1}
\chi_{nlm}(\textbf{r};\zeta)=S_n(\zeta)\, r^{n-1}\, e^{-\zeta r}\, Y_{lm}(\theta,\phi),
\end{align}
where $n$ and $l$ are both integers such that $n>l$, $(r,\theta,\phi)$ are the spherical coordinates of the given
centre, $S_n(\zeta)=(2\zeta)^{n+1/2}/\sqrt{(2n)!}$ is the (radial) normalisation constant, and $Y_{lm}$ are
spherical harmonics in the Condon-Shortley phase convention
\begin{align}
\label{ylm}
Y_{lm}(\hat{\bf r}) = \Omega_{lm} P_l^{|m|}(\cos \theta)\frac{e^{\dot{\imath}m\phi}}{\sqrt{2\pi}},
\end{align}
where $P_l^m$ are the (unnormalised) associated Legendre polynomials \cite{stegun72} and $\Omega_{lm}$ is the angular
normalisation constant
\begin{align}
\Omega_{lm} = \dot{\imath}^{m-|m|} \sqrt{\frac{2l+1}{2}\frac{(l-|m|)!}{(l+|m|)!}}.
\end{align}
Transformation to the real spherical harmonics, which are usually more convenient in calculations, can be performed with
standard relations.

Throughout the paper the electrons shall be denoted by $1,2,...$ and the nuclei by $a,b,...$. All interparticle
distances are shortly written 
as $r_{pq}$, \emph{e.g.}, the distance between the first electron and the nucleus $a$ is simply $r_{1a}$ \emph{etc.} (an
exception is the
internuclear distance for which the usual convention $R:=r_{ab}$ is adapted). Let us introduce the prolate ellipsoidal
coordinates, $(\xi_i,\eta_i,\phi_i)$, 
by means of the formulae
\begin{align}
\xi_i  = \frac{r_{ia}+r_{ib}}{R},\;\; \eta_i = \frac{r_{ia}-r_{ib}}{R},
\end{align}
where $i=1,2$, and $\phi_i$ are the corresponding azimuthal angles. The volume element becomes $d\textbf{r}_i =
\left(\frac{R}{2}\right)^3(\xi_i^2-\eta_i^2)\,d\xi_i\,d\eta_i\,d\phi_i$.
It is well known that the product of two Slater-type orbitals can be written in a closed-from in the prolate ellipsoidal
coordinate system as follows
\begin{align}
\label{trans12}
\begin{split}
&\left(\frac{R}{2}\right)^3(\xi_i^2-\eta_i^2)\chi_{n_al_am_a}^*(\textbf{r}_{ia};\zeta_a)\chi_{n_bl_bm_b}(\textbf{r}_{ib}
;\zeta_b)=\\
&\frac{K_{ab}}{2\pi}e^{-\alpha_i\xi_i-\beta_i\eta_i}
\left[(\xi_i^2-1)(1-\eta_i^2)\right]^{|M_i|/2}e^{\dot{\imath}M_i\phi}
\sum_{p,q=0}^\Gamma{\bf \large \Xi}_{pq} \xi_i^p\; \eta_i^q,
\end{split}
\end{align}
with $M=m_a-m_b$ and $\Gamma=l_a+l_b+2$. The new coefficients are defined as
$\alpha=\frac{R}{2}(\zeta_a+\zeta_b)$,
$\beta=\frac{R}{2}(\kappa_a\zeta_a+\kappa_b\zeta_b)$, $K_{ab}=S_{n_a}(\zeta_a) S_{n_b}(\zeta_b)\, \Omega_{l_am_a}
\Omega_{l_bm_b} \left(\frac{R}{2}\right)^{n_a+n_b+1}$. 
The quantity ${\bf \large \Xi}$ is a square matrix with some numerical coefficients which can be tabulated. Details of
this transformation are given in Refs. [\onlinecite{lesiuk2014b}]
(see also the references therein). In conclusion, any nonzero two-centre electron repulsion integral over STOs can be
written down as a finite linear combination of the following
generic integrals
\begin{align}
\label{gener}
\begin{split}
\mathcal{I}_{p_1q_1}^{p_2q_2}(\sigma) &= 
\int_1^\infty d\xi_1 \int_{-1}^{+1} d\eta_1 \int_0^{2\pi} d\phi_1 \int_1^\infty d\xi_2 \int_{-1}^{+1} d\eta_2
\int_0^{2\pi} d\phi_2 \,\xi_1^{p_1} \eta_1^{q_1} \xi_2^{p_2} \eta_2^{q_2} \\
&\times e^{-\alpha_1 \xi_1-\alpha_2 \xi_2 -\beta_1 \eta_1-\beta_2 \eta_2}
\left[(\xi_1^2-1)(1-\eta_1^2)(\xi_2^2-1)(1-\eta_2^2)\right]^{\sigma/2}\frac{1}{r_{12}}, 
\end{split}
\end{align}
where explicit notation for the nonlinear parameters has been suppressed for brevity. The values of $p_i$, $q_i$ and
$\sigma$ are restricted to non-negative integers.

Let us now introduce the Neumann expansion of the Coulomb interaction potential
\begin{align}
\begin{split}
\label{neumann}
\frac{1}{r_{12}}&=\frac{2}{R}\sum_{\mu=0}^\infty \sum_{\sigma=-\mu}^\mu (-1)^\sigma (2\mu+1) 
\left[ \frac{(\mu-|\sigma|)!}{(\mu+|\sigma|)!} \right]^2\\
&\times P_\mu^{|\sigma|}(\xi_<)Q_\mu^{|\sigma|}(\xi_>)P_\mu^{|\sigma|}(\eta_1)P_\mu^{|\sigma|}(\eta_2)
e^{\dot{\imath}\sigma(\phi_1-\phi_2)},
\end{split}
\end{align}
where $\xi_<=\min(\xi_1,\xi_2)$ and $\xi_>=\max(\xi_1,\xi_2)$, $P_l^m$ are defined in the same way as in Eq.
(\ref{ylm}), and $Q_l^m$ are the associated Legendre functions of the 
second kind. By plugging the above expansion into Eq. (\ref{gener}) and after a straightforward integration over the
angles one arrives at
\begin{align}
\label{neumann1}
\mathcal{I}_{p_1q_1}^{p_2q_2}(\sigma) = \frac{8}{R}(-1)^\sigma \sum_{\mu=\sigma}^\infty
(2\mu+1)W_\mu^\sigma(p_1,p_2,\alpha_1,\alpha_2)
i_\mu^\sigma(q_1,\beta_1)i_\mu^\sigma(q_2,\beta_2),
\end{align}
where the basic quantities for the integration over $\eta$ are
\begin{align}
\label{imus}
i_\mu^\sigma(q,\beta)=\frac{(-1)^\mu}{2}\frac{(\mu-\sigma)!}{(\mu+\sigma)!}
\int_{-1}^{+1} d\eta \,P_\mu^{|\sigma|}(\eta)
(1-\eta^2)^{\sigma/2}\eta^q\, e^{-\beta \eta},
\end{align}
and similarly for the $\xi$ integration
\begin{align}
\begin{split}
\label{bigw}
W_\mu^\sigma(p_1,p_2,\alpha_1,\alpha_2)&=w_\mu^\sigma(p_1,p_2,\alpha_1,\alpha_2)\\
&+w_\mu^\sigma(p_2,p_1,\alpha_2,\alpha_1),\\
\end{split}
\end{align}
\begin{align}
\label{smallw}
\begin{split}
w_\mu^\sigma(p_1,p_2,\alpha_1,\alpha_2)&=\int_1^\infty d\xi_1\, Q_\mu^{\sigma}(\xi_1)\,
(\xi_1^2-1)^{\sigma/2}\xi_1^{p_1}e^{-\alpha_1\xi_1}\\
&\times\int_1^{\xi_1}d\xi_2\, P_\mu^{\sigma}(\xi_2)\,
(\xi_2^2-1)^{\sigma/2}\xi_2^{p_2}e^{-\alpha_2\xi_2}.
\end{split}
\end{align}
In general, the expansion given by Eq. (\ref{neumann1}) is infinite and terminates only in the special case of vanishing
$\beta_1$ or $\beta_2$. Nonetheless, it is convergent for
any physically acceptable values of the nonlinear parameters, \emph{i.e.}, $\alpha_i>0$ and $|\beta_i|\leq\alpha_i$, but
the \emph{rate} of convergence depends crucially on the values
of $\beta_i$. A practical observation is that larger values of $\beta_i$ result in a slower convergence. Unfortunately,
in actual calculations one can expect some of the integrals to approach the extreme case $|\beta_i|=\alpha_i$. In such
situation several tens of terms may be
necessary to achieve convergence which significantly slows down the computations. Note parenthetically that the
convergence is somewhat slower for larger values of $\sigma$, but this effect is of 
secondary importance. 

Calculation of the integrals $i_\mu^\sigma(q,\beta)$ is not connected with any significant overhead, even if large
values of the parameters are necessary. Therefore, at present we see no reason to develop new methods for their
computation. The available techniques appear to be entirely
satisfactory and the recursive method put forward by Harris 
is particularly robust (see Ref. [\onlinecite{harris02}] for an extended survey). We shall concentrate on the most
difficult basic
quantities, \emph{i.e.}, the integrals $w_\mu^\sigma(p_1,p_2,\alpha_1,\alpha_2)$.
Let us recall the analytical formula derived by Maslen and Trefry (after simplifications due to Harris)
\begin{align}
\begin{split}
\label{wmugen}
&w_\mu^\sigma(p_1,p_2,\alpha_1,\alpha_2)=\Bigg[ \frac{(\mu+\sigma)!}{(\mu-\sigma)!} \Bigg]^2
\Bigg[ L_\mu^\sigma(p_1,\alpha_1) k_\mu^\sigma(p_2,\alpha_2)\\
&-\sum_s^\mu
\mathcal{A}_s^{\mu\sigma}\sum_{j=0}^{p_2+s}\frac{(p_2+s)!}{j!\alpha_2^{p_2+s}}L_\mu^\sigma(p_1+j,\alpha_1+\alpha_2)
\Bigg],
\end{split}
\end{align}
where
\begin{align}
\label{lmuint}
L_\mu^\sigma(p,\alpha)=\frac{(\mu-\sigma)!}{(\mu+\sigma)!}\int_1^\infty d\xi\,
Q_\mu^\sigma(\xi)\xi^p(\xi^2-1)^{\sigma/2} e^{-\alpha \xi},
\end{align}
and
\begin{align}
\label{kmu}
k_\mu^\sigma(p,\alpha)=\frac{(\mu-\sigma)!}{(\mu+\sigma)!}\int_1^\infty d\xi\,
P_\mu^\sigma(\xi)\xi^p(\xi^2-1)^{\sigma/2} e^{-\alpha \xi}.
\end{align}
The main goal of the present paper is to provide efficient and reliable methods for calculation of $W_\mu$ for large
values of $\mu$. The problem can be solved in two ways. The first
one is a direct attack by using the differential equation for $W_\mu$ derived in the previous paper. The second method
utilises Eq. (\ref{wmugen}) and reduces the problem to calculation
of $L_\mu^\sigma(p,\alpha)$ which appears to be more straightforward. In fact, large $\mu$ expansion of
$L_\mu^\sigma(p,\alpha)$ is expected to be significantly less complicated than the corresponding
one for $W_\mu$. However, there is an additional cost of using Eq. (\ref{wmugen}) which is absent in the first method
where $W_\mu$ are calculated directly. Let us also note in passing that the auxiliary integrals
$k_\mu^\sigma(p,\alpha)$, Eq.
(\ref{kmu}), can be computed efficiently with the available techniques and thus are not considered herein.

Throughout the paper we rely on two special functions, $E_n(z)$ and $a_n(z)$. They are defined in the Appendix A and
efficient methods of their computation are briefly discussed.

\section{Large-order expansion of $L_\mu^\sigma(p;\alpha)$}
\label{sec:expl}

\subsection{Initial reduction}
\label{subsec:expl1}

Let us recall two recursion relations which allow to 
simplify the problem significantly. They result directly from the properties of the Legendre functions and read
\begin{align}
\label{lgrowp}
L_\mu^0(p+1,\alpha) &= \frac{(\mu+1)L_{\mu+1}^0(p,\alpha)+\mu L_{\mu-1}^0(p,\alpha)}{2\mu+1},\\
\label{lgrows}
L_\mu^{\sigma+1}(p,\alpha) &= \frac{L_{\mu+1}^{\sigma}(p,\alpha)-L_{\mu-1}^{\sigma}(p,\alpha)}{2\mu+1}.
\end{align}
By means of these recursions the necessary integrals $L_\mu^\sigma(p;\alpha)$ can be efficiently computed starting with
$L_\mu(\alpha):=L_\mu^0(0;\alpha)$ only. Note that the above expressions require $L_\mu(\alpha)$ with even larger $\mu$
than
initially. In fact, they basically consist of increasing $p$ and $\sigma$ at cost of $\mu$. Therefore, the most
important task 
is to calculate the integrals $L_\mu(\alpha)$ for large $\mu$ with decent speed and precision. This is the main issue
considered in
the present section.

\subsection{Alternative integral representations of $L_\mu(\alpha)$}
\label{subsec:expl2}

Our derivation starts with the 
differential equation for $L_\mu(\alpha)$ which was established in Ref. [\onlinecite{lesiuk2014b}]
\begin{align}
\label{lmudiff}
\alpha^2 L_\mu''(\alpha)+2\alpha L_\mu'(\alpha)-\left[\mu(\mu+1)+\alpha^2\right]L_\mu(\alpha)=-e^{-\alpha},
\end{align}
where the prime denotes differentiation with respect to $\alpha$. Let us recall that the linearly independent
solutions
of the homogeneous differential equation are the well-known modified spherical Bessel functions \cite{stegun72},
$i_\mu(\alpha)$ and
$k_\mu(\alpha)$.
This suggests 
that the desired solution of the inhomogeneous equation has the following form
\begin{align}
\label{lmusten}
\mathcal{K}_\mu(\alpha)\, i_\mu(\alpha) + \mathcal{I}_\mu(\alpha)\, k_\mu(\alpha),
\end{align}
where $\mathcal{I}_\mu(\alpha)$ and $\mathcal{K}_\mu(\alpha)$ are some functions which are yet to be determined. Let us
additionally enforce
the constraint
\begin{align}
\label{lmuconstr}
\mathcal{K}_\mu'(\alpha)\, i_\mu(\alpha) + \mathcal{I}_\mu'(\alpha)\, k_\mu(\alpha) = 0,
\end{align}
valid strictly for every value of $\alpha>0$. Upon inserting the formula (\ref{lmusten}) into the differential equation
(\ref{lmudiff}) one 
obtains the following expression
\begin{align}
\mathcal{K}_\mu'(\alpha)\, i_\mu'(\alpha) + \mathcal{I}_\mu'(\alpha)\, k_\mu'(\alpha) = - \frac{e^{-\alpha}}{\alpha^2},
\end{align}
where we have taken advantage of the fact that $i_\mu(\alpha)$ and $k_\mu(\alpha)$ obey the homogeneous
differential equation. The above expression and the constraint (\ref{lmuconstr}) form the following system of
linear
equations
\begin{align}
\label{lmusys}
\left[
\begin{tabular}{cc}
 $i_\mu(\alpha)$  & $k_\mu(\alpha)$  \\
 $i'_\mu(\alpha)$ & $k'_\mu(\alpha)$ \\
\end{tabular} 
\right] 
\left[
\begin{tabular}{c}
 $\mathcal{K}_\mu'(\alpha)$ \\
 $\mathcal{I}_\mu'(\alpha)$ \\
\end{tabular} 
\right]=
\left[
\begin{tabular}{c}
 $-e^{-\alpha}/\alpha^2$ \\
 $0$ \\
\end{tabular}
\right]
\end{align}
Note that the determinant of the above $2\times 2$ matrix (the Wronskian) is equal to $-\frac{\pi}{2}\frac{1}{\alpha^2}$
which
is a direct consequence of the properties
of the Bessel functions\cite{stegun72}. The system of linear equations (\ref{lmusys}) can be solved right away,
\emph{e.g.}, with the Cramer's rule to give
\begin{align}
&\mathcal{K}_\mu'(\alpha) = +\frac{2}{\pi}\, e^{-\alpha}\, k_\mu(\alpha),\\
&\mathcal{I}_\mu'(\alpha) = -\frac{2}{\pi}\, e^{-\alpha}\, i_\mu(\alpha),
\end{align}
and after (indefinite) integration over $\alpha$ one arrives at
\begin{align}
\label{lmusolg}
\frac{2}{\pi}\, i_\mu(\alpha) \int d\alpha\, e^{-\alpha}\, k_\mu(\alpha) - \frac{2}{\pi}\, k_\mu(\alpha) \int d\alpha\,
e^{-\alpha}\, i_\mu(\alpha).
\end{align}
This expression is the general solution of the differential equation (\ref{lmudiff}). In order to find a particular
solution corresponding to the
integrals (\ref{lmuint}) we need to impose proper initial conditions. From Eq. (\ref{lmudiff}) one clearly sees that
$L_\mu(\alpha)$ vanish as $\alpha\rightarrow\infty$
for every $\mu$. Additionally, the results presented in Ref. [\onlinecite{lesiuk2014b}] indicate that $L_\mu(\alpha)$
vanish exponentially
quickly in this limit [as $e^{-\alpha}\log(\alpha)$ in the leading-order term].
This constitutes the first initial condition which we need to impose on the above general solution. The second initial
condition results from the small $\alpha$ behaviour of $L_\mu(\alpha)$
\begin{align}
\label{const2}
L_\mu^0(\mu,\alpha) =-\gamma_E \frac{\mu!}{(2\mu+1)!!}-\frac{\mu!}{(2\mu+1)!!} \log(2\alpha)+\widetilde{\mathcal{M}}_\mu
+ \mathcal{O}(\alpha),
\end{align}
where $\gamma_E$ is the Euler-Mascheroni constant, and $\widetilde{\mathcal{M}}_\mu$ are some numerical coefficients
independent of $\alpha$ (\emph{c.f.} the supplemental material to Ref. [\onlinecite{lesiuk2014b}]). An essential feature
of the above formula is
the logarithmic singularity for
small $\alpha$ which has to be reproduced by Eq. (\ref{lmusolg}).
 The most succinct formula which takes both initial conditions into account reads
\begin{align}
\label{lmunew}
&L_\mu(\alpha) = \mathcal{K}_\mu(\alpha)\, i_\mu(\alpha) + \mathcal{I}_\mu(\alpha)\, k_\mu(\alpha),\\
\label{lmunewa}
&\mathcal{K}_\mu(\alpha) = \frac{2}{\pi} \int_\alpha^\infty dz\, e^{-z}\, k_\mu(z),\\
\label{lmunewb}
&\mathcal{I}_\mu(\alpha) = \frac{2}{\pi} \int_0^\alpha dz\, e^{-z}\, i_\mu(z).
\end{align}
Clearly, the formula (\ref{lmunew}) is 
a new integral representation of the $L_\mu(\alpha)$ functions, alternative to the definition given by Eq.
(\ref{lmuint}).

At this point an extended comment is mandatory. One might be uncertain about the reason behind introduction of Eq.
(\ref{lmunew}). It is clearly more complicated than the initial 
definition, Eq. (\ref{lmuint}), and appears to give no computational or theoretical advantages. However, it turns out
that Eq. (\ref{lmuint}) is a very inconvenient starting point
for the present developments. Despite the large-order expansions of the Legendre functions, $Q_\mu$, are well-known
\cite{olver97,jones01,dunster04}, they are too complicated to be used for our purposes. 
A naive approach where a large-order expansion of $Q_\mu$ is inserted into Eq. (\ref{lmuint}) leads to
intractable integrals requiring a numerical solution. On the other hand,
Eq. (\ref{lmunew}) is formulated solely in terms of the modified spherical Bessel functions. This is advantageous, as
the large-order expansions of $i_\mu(\alpha)$ and $k_\mu(\alpha)$ are 
more compact and straightforward. In particular, we rely on the recent works of Sidi and Hoggan \cite{sidi11,hoggan14}
where an elegant
formulation has been given. For convenience of the readers we list the relevant formulae of Sidi and
Hoggan in the Appendix B, utilising our notation.

\subsection{Large-order expansion of $L_\mu(\alpha)$}
\label{subsec:expl3}

Having the integral representation (\ref{lmunew}) at hand, it becomes
straightforward to derive the large-order expansion of the pertinent
integrals $\mathcal{I}_\mu(\alpha)$ and $\mathcal{K}_\mu(\alpha)$. By inserting the integral representations
(\ref{imuasym}) and (\ref{kmuasym}) into Eqs. (\ref{lmunewa}) and (\ref{lmunewb}), respectively,
one obtains
\begin{align}
\label{lmuik}
&\mathcal{I}_\mu(\alpha) = \frac{2}{\sqrt{\pi}} \left(\frac{\alpha}{2}\right)^{\mu+1} \frac{1}{\Gamma(\mu+3/2)}
\sum_{m=0}^\infty \frac{\lambda_m^\mu(\alpha)}{(\mu+1/2)^m},\\
&\mathcal{K}_\mu(\alpha) = \frac{1}{\sqrt{\pi}} \left(\frac{2}{\alpha}\right)^\mu \Gamma(\mu+1/2)
\sum_{m=0}^\infty \frac{\Lambda_m^\mu(\alpha)}{(\mu+1/2)^m}(-1)^m,
\end{align}
after a simple exchange of variables. The coefficients in the expansions are obtained with help of Eq. (\ref{bm})
\begin{align}
\lambda_m^\mu(\alpha) &= \int_0^1 dt \, t^\mu\, b_m(\alpha t)\, e^{-\alpha t} = \sum_{k=1}^m (-1)^{m-k}
\frac{S_{mk}}{k!} (\alpha/2)^{2k} a_{\mu+2k}(\alpha),\\
\Lambda_m^\mu(\alpha) &= \int_1^\infty dt \, \frac{b_m(\alpha t)}{t^{\mu+1}}\, e^{-\alpha t} = \sum_{k=1}^m (-1)^{m-k}
\frac{S_{mk}}{k!} (\alpha/2)^{2k} E_{\mu-2k-1}(\alpha),
\end{align}
which is valid for $m>0$. In the special case of $m=0$ the corresponding results are
$\lambda_0^\mu(\alpha)=a_\mu(\alpha)$
and $\Lambda_0^\mu(\alpha)=E_{\mu+1}(\alpha)$. The modified spherical Bessel functions in Eq. (\ref{lmunew})
which multiply the integrals 
$\mathcal{I}_\mu(\alpha)$ and $\mathcal{K}_\mu(\alpha)$ can also be expanded with help of Eqs. (\ref{imuasym}) and
(\ref{kmuasym}). This leads to a product of two infinite series which
can finally be rewritten as
\begin{align}
\label{lmuasym}
&L_\mu(\alpha) = \frac{1}{2\mu+1} \sum_{s=0}^\infty \frac{c_s^\mu(\alpha)}{(\mu+1/2)^s},\\
&c_s^\mu(\alpha) = \sum_{l=0}^s b_{s-l}(\alpha) (-1)^l \Big[ \Lambda_l^\mu(\alpha) + (-1)^s \lambda_l^\mu(\alpha)
\Big].
\end{align}
A short remark on the mathematical nature of the above expansion is necessary. Note that Eq. (\ref{lmuasym}) would not
be classified as an asymptotic expansion by some authors because
the coefficients $c_s^\mu$ are explicitly $\mu$-dependent. That is why we prefer to use the term large-order expansion.
We verified that Eq. (\ref{lmuasym}) is capable of providing
arbitrarily accurate results if only the value of $\mu$ is sufficiently large. Thus, from the pragmatic point of view,
Eq. (\ref{lmuasym}) gives an effective method to calculate the values
of $L_\mu(\alpha)$ for large $\mu$ where other techniques run out of steam.

From the point of view of some developments it is useful to analyse in details the first term of the expansion
(\ref{lmuasym}). One easily arrives at
\begin{align}
L_\mu(\alpha) = \frac{1}{2\mu+1} \left[ E_{\mu+1}(\alpha) + a_\mu(\alpha) \right] + \mbox{h.o.}
\end{align}
Additionally, if the large $\mu$ asymptotic formulae for $E_{\mu+1}(\alpha)$ and $a_\mu(\alpha)$ are used, Eqs.
(\ref{enasym}) and (\ref{anasym}), some simplifications occur and one finds
\begin{align}
L_\mu(\alpha) = \frac{e^{-\alpha}}{2\mu+1} \left[ \frac{1}{\mu-\alpha} + \frac{1}{\mu+\alpha} \right] +
\mathcal{O}\left( \frac{1}{\mu^3} \right),
\end{align}
provided that $\mu>\alpha$.

\subsection{Numerical tests and examples}
\label{subsec:lnum}

It is now mandatory to verify how the new formula (\ref{lmuasym}) works in practice. In Table \ref{table1} we present
results of
some exemplary calculations of $L_\mu(\alpha)$ with help of the new formula, Eq. (\ref{lmuasym}).
Different values of $\alpha$ and $\mu$ are tested to find the actual range of applicability. Additionally, the number
of terms in the infinite expansion (\ref{lmuasym}) necessary to reach the maximal possible
precision was listed in each case. A more detailed inspection of Table \ref{table1} reveals some general conclusions
about the range of
the parameters where Eq. (\ref{lmuasym}) gives sufficiently accurate results. One sees
that the convergence of the infinite summation in Eq. (\ref{lmuasym}) is excellent for small or moderate values of
$\alpha$. Unfortunately, it deteriorates quickly when the values of $\alpha$ and $\mu$ approach each 
other. In the case when $\alpha>\mu$ no useful information about $L_\mu(\alpha)$ can be obtained with help of Eq.
(\ref{lmuasym}). However, this is not a reason for a major concern. In fact, the 
large $\alpha$ expansion of $L_\mu(\alpha)$ was given in Ref. [\onlinecite{lesiuk2014b}] and it works reasonably well
for both small and large
values of $\mu$. We
conclude that Eq. (\ref{lmuasym}) is a preferred computational technique when $\mu$ is large and $\alpha$
is small or moderate at the same time.

\section{Large-order expansion of $W_\mu^\sigma(p_1,p_2;\alpha_1,\alpha_2)$}
\label{sec:expw}

\subsection{Initial reduction}
\label{subsec:expw1}

Let us reduce the number of independent parameters in the integrals $W_\mu$ by using two convenient formulae. 
The first one is the remainder in the recursive method proposed by Kotani \cite{kotani55}
\begin{align}
\label{wmugrows}
\begin{split}
&W_\mu^{\sigma+1}(p_1,p_2,\alpha_1,\alpha_2)=
\frac{(\mu-\sigma)(\mu-\sigma+1)^2}{2\mu+1}W_{\mu+1}^\sigma(p_1,p_2,\alpha_1,\alpha_2)-(\mu-\sigma)(\mu+\sigma+1)\\
&\times W_\mu^\sigma(p_1+1,p_2+1,\alpha_1,\alpha_2)+\frac{(\mu+\sigma+1)(\mu+\sigma)^2}{2\mu+1}
W_{\mu-1}^\sigma(p_1,p_2,\alpha_1,\alpha_2).
\end{split}
\end{align}
This expression is numerically stable for a wide range of the parameters values. As a result,
it constitutes a reliable method for computation of $W_\mu^\sigma(p_1,p_2,\alpha_1,\alpha_2)$ from
the integrals with $\sigma=0$. Additionally, the values of $p_2$ can be increased by differentiation
\begin{align}
\label{wmugrowp2}
W_\mu^0(p_1,p_2,\alpha_1,\alpha_2) = (-1)^{p_2} \frac{\partial^{p_2}}{\partial \alpha_2^{p_2}}
W_\mu^0(p_1,0,\alpha_1,\alpha_2).
\end{align}
Further in the article, we consider the large $\mu$ expansion of the basic integrals $W_\mu^0(p_1,0,\alpha_1,\alpha_2)$.
Note that differentiation with respect to $\alpha_1$ could be used to increase the value of $p_1$, but this
approach is not particularly advantageous in the present context. For convenience, we introduce the following shorthand
notation, $W_{\mu}(p;\alpha_1,\alpha_2)=W_\mu^0(p,0,\alpha_1,\alpha_2)$.

\subsection{Alternative integral representations of $W_\mu^0(p_1,0,\alpha_1,\alpha_2)$}
\label{subsec:expw2}

Let us recall the differential equation for $W_{\mu}(p;\alpha_1,\alpha_2)$ obtained in Ref. [\onlinecite{lesiuk2014b}]
\begin{align}
\label{bigwdiff}
\begin{split}
&\alpha_2^2\,\frac{\partial^2}{\partial\alpha_2^2}\,W_\mu(p;\alpha_1,\alpha_2)+2\alpha_2\,\frac{\partial}{
\partial\alpha_2}\,W_\mu(p;\alpha_1,\alpha_2)\,+\\
&-\Big[\mu(\mu+1)+\alpha_2^2\Big]W_\mu(p;\alpha_1,\alpha_2)=-E_{-p}(\alpha_1+\alpha_2),
\end{split}
\end{align}
which provides the starting point for our derivation. Note that the solutions of the homogeneous equation
are well-known and are the same as for Eq. (\ref{lmudiff}). Therefore, the solution can be written in the
form analogous to Eq. (\ref{lmusten}) and the derivation follows along a very similar line as for $L_\mu$. There is no
need to repeat details of the derivation and we present only the final result
\begin{align}
\label{wmunew}
\begin{split}
&W_{\mu}(p;\alpha_1,\alpha_2) = 
\mathcal{K}_\mu(p;\alpha_1,\alpha_2)\, i_\mu(\alpha_2) + 
\mathcal{I}_\mu(p;\alpha_1,\alpha_2)\, k_\mu(\alpha_2),\\
&\mathcal{K}_\mu(p;\alpha_1,\alpha_2) = \frac{2}{\pi} \int_{\alpha_2}^\infty dz\, k_\mu(z)\, E_{-p}(\alpha_1+z),\\
&\mathcal{I}_\mu(p;\alpha_1,\alpha_2) = \frac{2}{\pi} \int_0^{\alpha_2} dz\, i_\mu(z)\, E_{-p}(\alpha_1+z),
\end{split}
\end{align}
by imposing proper initial conditions (\emph{c.f.} Ref. [\onlinecite{lesiuk2014b}]). Note that the basic integrals were
expressed though the
modified spherical Bessel functions, in analogy with $L_\mu$ functions considered before. Clearly, Eqs. (\ref{wmunew})
may
be useful on their own (\emph{e.g.} evaluation by a numerical integration), but in the present paper we concentrate
solely on the large 
$\mu$ expansion of $W_{\mu}(p;\alpha_1,\alpha_2)$.

\subsection{Large-order expansion of $W_\mu^0(p_1,0,\alpha_1,\alpha_2)$}
\label{subsec:expw3}

Let us now insert the asymptotic expansions of $i_\mu(\alpha)$ and $k_\mu(\alpha)$, Eqs. (\ref{imuasym}) and
(\ref{kmuasym}), into the integral representation (\ref{wmunew}). After straightforward rearrangements one arrives at
\begin{align}
\label{wmuik}
&\mathcal{I}_\mu(p;\alpha_1,\alpha_2) = \frac{2}{\sqrt{\pi}} \left(\frac{\alpha_2}{2}\right)^{\mu+1} \frac{1}{\Gamma(\mu+3/2)}
\sum_{m=0}^\infty \frac{\tau_m^{\mu p}(\alpha_1,\alpha_2)}{(\mu+1/2)^m},\\
&\mathcal{K}_\mu(p;\alpha_1,\alpha_2) = \frac{1}{\sqrt{\pi}} \left(\frac{2}{\alpha_2}\right)^\mu \Gamma(\mu+1/2)
\sum_{m=0}^\infty \frac{T_m^{\mu p}(\alpha_1,\alpha_2)}{(\mu+1/2)^m}(-1)^m,
\end{align}
which is an analogue of Eqs. (\ref{lmuik}). The analytical formulae for the coefficients are obtained by recalling Eq.
(\ref{bm})
\begin{align}
\label{tau}
\begin{split}
\tau_m^{\mu p}(\alpha_1,\alpha_2) &= \int_0^1 dt \, t^\mu\, b_m(\alpha_2 t)\, E_{-p}(\alpha_1+\alpha_2 t) \\
&= \sum_{k=1}^m (-1)^{m-k} \frac{S_{mk}}{k!}\, (\alpha_2/2)^{2k}\, \omega_{\mu+2k,p}(\alpha_1,\alpha_2),
\end{split}
\end{align}
\begin{align}
\label{Tau}
\begin{split}
T_m^{\mu p}(\alpha_1,\alpha_2) &= \int_1^\infty dt \, \frac{b_m(\alpha t)}{t^{\mu+1}}\, E_{-p}(\alpha_1+\alpha_2 t) \\
&= \sum_{k=1}^m (-1)^{m-k}\frac{S_{mk}}{k!}\, (\alpha_2/2)^{2k}\, \Omega_{2k-\mu-1,p}(\alpha_1,\alpha_2),
\end{split}
\end{align}
for $m>1$, and $\tau_m^{\mu p}(\alpha_1,\alpha_2) = \omega_{\mu p}(\alpha_1,\alpha_2)$, $T_m^{\mu p}(\alpha_1,\alpha_2)
= \Omega_{-\mu-1,p}(\alpha_1,\alpha_2)$. The basic integrals are defined as
\begin{align}
\label{omegabasic}
\omega_{np}(\alpha_1,\alpha_2) &= \int_0^1 dt\, t^n E_{-p}(\alpha_1+\alpha_2 t),\\
\label{Omegabasic}
\Omega_{np}(\alpha_1,\alpha_2) &= \int_1^\infty dt\,t^n E_{-p}(\alpha_1+\alpha_2 t).
\end{align}
Note that evaluation of Eqs. (\ref{tau}) and (\ref{Tau}) requires $\omega_{np}$ with $n>0$, $p>0$, but
in the case of $\Omega_{np}$ the values of $n$ can be negative. Additionally, the first argument ($n$) in $\omega_{np}$
is always larger than $\mu$ [\emph{c.f.} Eq. (\ref{tau})]. The present method is intended to be used for large $\mu$ and
we concentrate on evaluation of $\omega_{np}$ with large $n$. Unfortunately, for the integrals
$\Omega_{np}$ such simplifications do not occur and more general methods are required. Calculation of the basic
integrals is discussed in the next section, with a considerable emphasis on the numerical stability.

Finally, one combines the asymptotic expansions (\ref{wmuik}) with the initial formula, Eq. (\ref{wmunew}), and after
some rearrangements the following expression is obtained
\begin{align}
\label{wmuasym}
 W_{\mu}(p;\alpha_1,\alpha_2) = \frac{1}{2\mu+1} \sum_{s=0}^\infty \frac{d_s^{\mu p}}{(\mu+1/2)^s},
\end{align}
where we have suppressed the notation for the nonlinear parameters, and
\begin{align}
d_s^{\mu p} = \sum_{l=0}^s b_{s-l}(\alpha_2) (-1)^l \Big[ T_l^{\mu p}(\alpha_1,\alpha_2) + (-1)^s \tau_l^{\mu
p}(\alpha_1,\alpha_2) \Big],
\end{align}
which constitutes the main result of the present section.

\subsection{Calculation of the basic integrals}
\label{subsec:expw4}

Let us begin with calculation of the integrals $\omega_{np}$. Integration of Eq. (\ref{omegabasic}) by parts leads to
the following recursion
\begin{align}
 \omega_{np}(\alpha_1,\alpha_2) = \frac{1}{n+1}\,E_{-p}(\alpha_1+\alpha_2) + \frac{\alpha_2}{n+1}\,
\omega_{n+1,p+1}(\alpha_1+\alpha_2 t).
\end{align}
In principle, the above relation can be used to calculate the values of $\omega_{np}$ by downward recursion, starting at
some large $n$ with an arbitrary value. However, the main drawbacks of this approach are difficulties in controlling
the error and choice of the starting point. Therefore, we propose to iterate this recursion analytically $N$ times
which gives
\begin{align}
&\omega_{np}(\alpha_1,\alpha_2) = n! \sum_{k=0}^N \frac{\alpha_2^k}{(n+k+1)!}\, E_{-p-k}(\alpha_1+\alpha_2) + R_N, \\
&R_N = \omega_{n+N+1,p+N+1}(\alpha_1,\alpha_2).
\end{align}
Note that the above expression is formally exact for each $N$. Additionally, when $n$ is large the terms in the above
sum vanish very quickly and large values of $N$ give very small contributions to the total value of the integral.
Similarly, the remainder $R_N$ vanishes fast with increasing $N$. To estimate in advance the required values of $N$ we
establish approximate upper bounds for the values of $R_N$ (note that $R_N$ is positive by
definition). Let us first insert the integral representation of $E_{-p}$, Eq. (\ref{En}), into Eq. (\ref{omegabasic})
and reverse the order of integrations. One arrives at the alternative integral representation of the remainder
\begin{align}
 R_N = \int_1^\infty dz\, z^{p+N+1} e^{-\alpha_1 z} \int_0^1 dt\, t^{n+N+1} e^{-\alpha_2 t z},
\end{align}
which is strictly bounded from above by
\begin{align}
 R_N \leq \int_1^\infty dz\, z^{p+N+1} e^{-\alpha_1 z} \int_0^1 dt\, t^{n+N+1} =
\frac{1}{n+N+2}\,E_{-p-N-1}(\alpha_1).
\end{align}
Additionally, one can verify that $E_{-p-N-1}(\alpha_1)$ is bounded from above by $(p+N+1)!/\alpha_1^{p+N+1}$. This
finally gives the estimation
\begin{align}
 R_N < \frac{n!}{\alpha_1^{p+1}} \left( \frac{\alpha_2}{\alpha_1} \right)^{N+1} \frac{(p+N+1)!}{(n+N+2)!}.
\end{align}

Passing to the integrals $\Omega_{np}$, the optimal algorithm depends on the sign of $n$. Similarly as before, by
inserting the integral representation of $E_{-p}$, Eq. (\ref{En}), into Eq. (\ref{Omegabasic}) and exchanging the order
of integrations
\begin{align}
 \Omega_{np}(\alpha_1,\alpha_2) = \int_1^\infty dz\, z^{p}\, e^{-\alpha_1 z} \int_1^\infty dt\, t^{n}\, e^{-\alpha_2
t z}= \int_1^\infty dz\, z^{p}\, e^{-\alpha_1 z}\, E_{-n}(\alpha_2 z).
\end{align}
When the values of $n$ are positive one can use the recursive relation (\ref{enrec}) which gives
\begin{align}
 \Omega_{np}(\alpha_1,\alpha_2) = \frac{n}{\alpha_2}\, \Omega_{n-1,p-1}(\alpha_1,\alpha_2)+
\frac{1}{\alpha_2}E_{-p+1}(\alpha_1+\alpha_2).
\end{align}
This recursion relation is completely stable when carried out in the upward direction along the ``diagonal'' lines.
However, it is not self-starting and requires values of $\Omega_{n0}$ and $\Omega_{0p}$ to initiate. Analytical
expression for the latter is fairly obvious, $\Omega_{0p}=E_{-p+1}(\alpha_1+\alpha_2)$, and calculation of the former
is based on the following relation
\begin{align}
 \Omega_{n0}(\alpha_1,\alpha_2) = \frac{e^{-\alpha_1}}{\alpha_2} E_{-n+1}(\alpha_2) -
\frac{\alpha_1}{\alpha_2}\Omega_{n-1,0}(\alpha_1,\alpha_2),
\end{align}
starting with $\Omega_{00}=E_1(\alpha_1+\alpha_2)$. This recursion is stable provided that the value of $\alpha_2$ is
moderate or large. If $\alpha_2$ is small the following series expansion is used
\begin{align}
 \Omega_{n0}(\alpha_1,\alpha_2) = \frac{n!}{\alpha_2^{n+1}} E_{n+1}(\alpha_1)-\sum_{k=0}^\infty
\frac{(-\alpha_2)^k}{k!}\frac{E_{-k}(\alpha_1)}{n+k+1},
\end{align}
which can be derived by using elementary methods. Similarly as before one can verify that the absolute value of each
term in the above sum is bounded by $\alpha_2^k/\alpha_1^{k+1}$ which can be used to estimate the convergence rate.

Finally, let us discuss calculation of $\Omega_{np}$ for negative values of $n$. The following recursion can be
derived with help of Eqs. (\ref{Omegabasic}) and (\ref{enrec})
\begin{align}
 \Omega_{np}(\alpha_1,\alpha_2) = \frac{p}{\alpha_2} \Omega_{n-1,p-1}(\alpha_1,\alpha_2) - \frac{\alpha_1}{\alpha_2}
\Omega_{n-1,p}(\alpha_1,\alpha_2) + \frac{e^{-\alpha_1}}{\alpha_2} E_{1-n}(\alpha_2),
\end{align}
which can be used to increase $p$ at cost of $n$. This recursive relation introduces some instabilities into
the calculation, but this fact is not significant as the values of $p$ rarely exceed 10. To initiate the above recursion
one requires the values of $\Omega_{n0}$. Similarly as before, the following expression is straightforward to derive
\begin{align}
 \Omega_{n0}(\alpha_1,\alpha_2) = \frac{e^{-\alpha_1}}{\alpha_2} E_{1-n}(\alpha_2) - \frac{\alpha_1}{\alpha_2}
\Omega_{n-1,0}(\alpha_1,\alpha_2).
\end{align}
This recursion relation is carried out downward, starting with $\Omega_{-N,0}$ at some large $N$. This completes the
formalism of calculation of the basic integrals.

\subsection{Numerical tests and examples}
\label{subsec:wnum}

In Tables \ref{table2} and \ref{table3} we present results of exemplary calculations of the
$W_{\mu}(p;\alpha_1,\alpha_2)$ functions with Eq. (\ref{wmuasym}) and comparison with the exact results. One can see
that the method based on Eq. (\ref{wmuasym}) converges in at most few tens of terms provided that $\alpha_1$ and
$\alpha_2$ are both small or moderate. In fact, for small values of the nonlinear parameters we managed to obtain the
convergence even for $\mu$ as small as 10 which shows the potential of the method. Unfortunately, when $\alpha_1$ and
$\alpha_2$ are both large (larger than 50, say) the series (\ref{wmuasym}) have an oscillatory behaviour and no
convergence was achieved after summing 200 terms. However,
in the regime of large $\alpha_1$ and $\alpha_2$ one can resort to different techniques \emph{e.g.} asymptotic
expansions presented in Ref. [\onlinecite{lesiuk2014b}]. Moreover, when $\alpha_1$ and $\alpha_2$ are simultaneously
large the resulting integrals are expected to be very small and they are likely to be negligible. Table
\ref{table2} lists the results for $p=0$ whilst the corresponding values for $p=8$ are given in Table \ref{table3}. A
more detailed comparison reveals that larger values of $p$ are connected with slower convergence of the series
(\ref{wmuasym}), but the range of applicability remains roughly the same.

\section{Conclusions}
\label{sec:conclusions}

We have presented a new systematic approach to the calculation of basic quantities appearing in the ellipsoidal
expansion of the two-electron integrals over Slater-type orbitals. Large-order ($\mu$) expansions of the functions
$L_\mu$ and $W_\mu$ have been given and their accuracy and range of applicability has been determined numerically. The
new method allows to calculate higher-order terms of the Neumann expansion with a significantly reduced computational
cost. As a result, this is a step towards reduction of the gap in computational timings between STOs and GTOs.
Moreover, the presented expressions may be useful in mathematical studies of convergence of the Neumann series and
rational design of convergence acceleration techniques.

\begin{acknowledgments}
This work was supported by the Polish Ministry of Science and Higher Education through the project
\textit{``Diamentowy Grant''}, number DI2011 012041.
\end{acknowledgments}

\appendix

\section{Auxiliary integrals}

Virtually all final working formulae obtained in the present paper are given in terms of the basic integrals
$E_n(z)$ and $a_n(z)$. They are defined through the integral representations
\begin{align}
\label{En}
&E_n(z) = \int_1^\infty dt\, \frac{e^{-z t}}{t^n}, \\
\label{an}
&a_n(z) = \int_0^1 dt\, t^n e^{-z t},
\end{align}
where $n$ is an arbitrary integer in the former and a nonnegative integer in the latter. Calculation of $a_n$ is most
easily carried out with help of the Miller algorithm \cite{gautschi67} as discussed by Harris \cite{harris02}. The
integral $E_n$ is usually called the generalised exponential integral. Computation of $E_n$ differs
depending on the sign of $n$. For a negative integer $n$ the following recursion is completely stable in the upward
direction
\begin{align}
\label{enrec}
E_n(z) = -\frac{n}{z} E_{n-1}(z)+\frac{e^{-z}}{z}.
\end{align}
For positive $n$ and $z<1$ one uses the series expansion
\begin{align}
E_n(z) = \frac{(-z)^{n-1}}{(n-1)!}\left[\Psi(n)-\log z\right]
-\sum_{\substack{k=0 \\ k\neq n-1}}^\infty \frac{(-z)^k}{k!\,(1-n+k)},
\end{align}
where $\Psi(n)$ is the digamma function at integer argument.
The above infinite summations converge to the machine precision in, at most, few tens of terms. Finally, for positive
$n$ and $z>1$ the continued fraction (CF) formula can be applied
\begin{align}
\label{encf}
E_n(z) = e^{-z} \left( \frac{1}{z+}\frac{p}{1+}\frac{1}{z+}\frac{p+1}{1+}\frac{2}{z+}\cdots \right).
\end{align}
To evaluate the CF one can use the Lentz algorithm \cite{stegun72}. The only inconvenience is that consecutive
numerators and
denominators in the 
Lentz scheme grow very quickly with the number of terms retained in Eq. (\ref{encf}). Therefore, it is necessary to
rescale them
from time to time by a small number to avoid
numerical overflows. Let us also recall the leading terms of the large-order asymptotic expansions for $E_n(z)$ and
$a_n(z)$ which read
\begin{align}
\label{enasym}
E_n(z) = \frac{e^{-z}}{n+z} + \mathcal{O}\left(\frac{1}{n^2}\right),\\
\label{anasym}
a_n(z) = \frac{e^{-z}}{n-z} + \mathcal{O}\left(\frac{1}{n^2}\right),
\end{align}
where $n>z$.

\section{Large-order asymptotic formulae for $i_\mu(\alpha)$ and $k_\mu(\alpha)$}

According to the work of Sidi \emph{et al.} \cite{sidi11,hoggan14} the modified spherical Bessel functions posses the
following large-order expansions
\begin{align}
\label{imuasym}
i_\mu(z) &= \frac{\sqrt{\pi}}{2}\frac{(z/2)^\mu}{\Gamma(\mu+\frac{3}{2})}\sum_{m=0}^\infty \frac{b_m(z)}{(\mu+1/2)^m},\\
\label{kmuasym}
k_\mu(z) &= \frac{\sqrt{\pi}}{4}\frac{\Gamma(\mu+\frac{1}{2})}{(z/2)^{\mu+1}}\sum_{m=0}^\infty
\frac{b_m(z)}{(\mu+1/2)^m}(-1)^m,
\end{align}
as $\mu\rightarrow\infty$ at a fixed $z$, where
\begin{align}
\label{bm}
b_m(z)=\sum_{k=1}^m (-1)^{m-k} \frac{S_{mk}}{k!} (z/2)^{2k},
\end{align}
for $m>0$ and $b_0(z)=1$. The quantities $S_{mk}$ in the above expression are the Stirling numbers of the
second kind \cite{stegun72} defined recursively as
\begin{align}
&S_{m0}=\delta_{m0},\;\;\; S_{m1}=1,\;\;\; S_{mm}=1,\;\;\;\\
&S_{mk}=S_{m-1,k-1}+k S_{m-1,k},
\end{align}
and we additionally adapt the convention $S_{mk}=0$ for $m<k$ or $m<0$.

\newpage
\LTcapwidth=\textwidth
\renewcommand{\tabcolsep}{12.0pt}
\begin{longtable}{cccc}
\caption{\label{table1} The functions $L_\mu(\alpha)$ calculated for some representative values of
$\alpha$ and $\mu$. \emph{Exact} denotes values calculated using explicit expressions (\emph{c.f.} Ref.
[\onlinecite{lesiuk2014b}]) in extended
arithmetic precision with the \textsc{Mathematica} package (all digits shown are correct). \emph{Large-order expansion}
column shows results of calculations with Eq. (\ref{wmuasym}) in the double precision arithmetic. \emph{Convergence}
denotes a number of terms in Eq. (\ref{wmuasym}) required to converge the summation to relative precision of
$2\cdot10^{-16}$. The symbol [k] denotes the powers of 10, 10$^k$.} \\
\hline\hline
$\mu$ & exact & large-order expansion & convergence \\
\hline
\multicolumn{4}{c}{$\alpha=0.1$} \\
\hline
30 & 9.72 733 864 877 071 [$-$04] & 9.72 733 864 877 071 [$-$04] & 9\\
40 & 5.51 662 783 117 224 [$-$04] & 5.51 662 783 117 224 [$-$04] & 9\\
50 & 3.54 810 355 237 372 [$-$04] & 3.54 810 355 237 372 [$-$04] & 7\\
60 & 2.47 209 822 882 328 [$-$04] & 2.47 209 822 882 328 [$-$04] & 7\\
\hline
\multicolumn{4}{c}{$\alpha=1.0$} \\
\hline
30 & 3.94 720 438 208 518 [$-$04] & 3.94 720 438 208 518 [$-$04] & 13\\
40 & 2.24 043 509 438 319 [$-$04] & 2.24 043 509 438 319 [$-$04] & 11\\
50 & 1.44 153 386 177 520 [$-$04] & 1.44 153 386 177 520 [$-$04] & 9\\
60 & 1.00 458 613 132 488 [$-$04] & 1.00 458 613 132 488 [$-$04] & 9\\
\hline
\multicolumn{4}{c}{$\alpha=10.0$} \\
\hline
30 & 4.78 078 398 572 794 [$-$08] & 4.78 078 398 572 794 [$-$08] & 29\\
40 & 2.73 528 592 642 051 [$-$08] & 2.73 528 592 642 051 [$-$08] & 21\\
50 & 1.76 662 929 639 839 [$-$08] & 1.76 662 929 639 839 [$-$08] & 19\\
60 & 1.23 372 624 613 915 [$-$08] & 1.23 372 624 613 915 [$-$08] & 19\\
\hline
\multicolumn{4}{c}{$\alpha=30.0$} \\
\hline
30 & 9.48 264 212 820 654 [$-$17] & divergence & $-$ \\
40 & 5.51 072 668 384 366 [$-$17] & divergence & $-$ \\
50 & 3.58 705 498 786 413 [$-$17] & 3.58 705 498 786 412 [$-$17] & 51\\
60 & 2.51 610 451 657 133 [$-$17] & 2.51 610 451 657 133 [$-$17] & 41\\
\hline\hline
\end{longtable}
\newpage

\begin{center}
\begin{longtable}{cccc}
\caption[aa]{The functions $W_\mu(p;\alpha_1,\alpha_2)$ calculated for some representative
values of the parameters. \emph{Exact} denotes values calculated using explicit expressions (\emph{c.f.} Ref.
[\onlinecite{lesiuk2014b}]) in
extended arithmetic precision with the \textsc{Mathematica} package (all digits shown are correct). \emph{Large-order
expansion} column shows results of calculations with Eq. (\ref{lmuasym}) in the double precision arithmetic.
\emph{Convergence} denotes a number of terms in Eq. (\ref{lmuasym}) required to converge the summation to relative
precision of $2\cdot10^{-16}$. The symbol [k] denotes the powers of 10, 10$^k$. } \label{table2} \\
\hline\hline
$\mu$ & exact & large-order expansion & convergence \\
\hline \endfirsthead
\multicolumn{4}{l}%
{{\tablename\ \thetable{}: Continued from the previous page.}} \\
\endhead
\multicolumn{4}{r}{{Continued on the next page}} \\
\endfoot
\hline \hline
\endlastfoot
\multicolumn{4}{c}{$\alpha_1=1.0$, $\alpha_2=1.0$, $p=0$} \\
\hline
30 & 7.26 438 420 525 738 [$-$05] & 7.26 438 420 525 741 [$-$05] & 13\\
40 & 4.12 230 721 703 987 [$-$05] & 4.12 230 721 703 989 [$-$05] & 11\\
50 & 2.65 207 347 050 162 [$-$05] & 2.65 207 347 050 163 [$-$05] & 9\\
60 & 1.84 808 542 738 505 [$-$05] & 1.84 808 542 738 506 [$-$05] & 7\\
\hline
\multicolumn{4}{c}{$\alpha_1=1.0$, $\alpha_2=5.0$, $p=0$} \\
\hline
30 & 4.43 294 559 944 128 [$-$07] & 4.43 294 559 944 129 [$-$07] & 19\\
40 & 2.51 607 499 568 558 [$-$07] & 2.51 607 499 568 559 [$-$07] & 16\\
50 & 1.61 886 531 395 923 [$-$07] & 1.61 886 531 395 924 [$-$07] & 12\\
60 & 1.12 815 856 115 223 [$-$07] & 1.12 815 856 115 224 [$-$07] & 9\\
\hline
\multicolumn{4}{c}{$\alpha_1=1.0$, $\alpha_2=10.0$, $p=0$} \\
\hline
30 & 1.62 914 653 440 044 [$-$10] & 1.62 914 653 440 043 [$-$10] & 29\\
40 & 9.24 696 775 490 950 [$-$10] & 9.24 696 775 490 946 [$-$10] & 21\\
50 & 5.94 963 334 102 345 [$-$10] & 5.94 963 334 102 343 [$-$10] & 19\\
60 & 4.14 621 345 113 117 [$-$10] & 4.14 621 345 113 115 [$-$10] & 19\\
\hline
\multicolumn{4}{c}{$\alpha_1=1.0$, $\alpha_2=50.0$, $p=0$} \\
\hline
30 & 1.49 278 026 722 289 [$-$27] & divergence & -- \\
40 & 8.47 302 548 103 673 [$-$28] & divergence & -- \\
50 & 5.45 168 822 712 533 [$-$28] & divergence & -- \\
60 & 3.79 921 108 103 191 [$-$28] & 3.79 921 108 103 191 [$-$28] & 157\\
\hline
\multicolumn{4}{c}{$\alpha_1=5.0$, $\alpha_2=5.0$, $p=0$} \\
\hline
30 & 4.85 315 094 815 931 [$-$09] & 4.85 315 094 815 933 [$-$09] & 21\\
40 & 2.75 906 112 497 927 [$-$09] & 2.75 906 112 497 928 [$-$09] & 17\\
50 & 1.77 656 445 608 724 [$-$09] & 1.77 656 445 608 724 [$-$09] & 15\\
60 & 1.23 857 631 621 439 [$-$09] & 1.23 857 631 621 440 [$-$09] & 13\\
\hline
\multicolumn{4}{c}{$\alpha_1=5.0$, $\alpha_2=10.0$, $p=0$} \\
\hline
30 & 2.17 631 353 217 224 [$-$11] & 2.17 631 353 217 225 [$-$11] & 29\\
40 & 1.23 815 626 851 864 [$-$11] & 1.23 815 626 851 865 [$-$11] & 21\\
50 & 7.97 525 834 542 020 [$-$12] & 7.97 525 834 542 023 [$-$12] & 19\\
60 & 5.56 120 404 111 395 [$-$12] & 5.56 120 404 111 398 [$-$12] & 19\\
\hline
\multicolumn{4}{c}{$\alpha_1=5.0$, $\alpha_2=50.0$, $p=0$} \\
\hline
30 & 2.51 589 622 982 829 [$-$29] & divergence & -- \\
40 & 1.43 273 896 436 017 [$-$29] & divergence & -- \\
50 & 9.23 286 988 543 448 [$-$30] & divergence & --\\
60 & 6.43 979 747 051 030 [$-$30] & 6.43 979 747 051 030 [$-$30] & 189\\
\hline
\multicolumn{4}{c}{$\alpha_1=10.0$, $\alpha_2=10.0$, $p=0$} \\
\hline
30 & 1.09 589 723 489 100 [$-$13] & 1.09 589 723 489 100 [$-$13] & 29\\
40 & 6.24 425 513 575 906 [$-$14] & 6.24 425 513 575 909 [$-$14] & 21\\
50 & 4.02 496 564 881 221 [$-$14] & 4.02 496 564 881 222 [$-$17] & 19\\
60 & 2.80 774 946 310 143 [$-$14] & 2.80 774 946 310 144 [$-$17] & 19\\
\hline
\multicolumn{4}{c}{$\alpha_1=10.0$, $\alpha_2=50.0$, $p=0$} \\
\hline
30 & 1.54 163 263 060 580 [$-$31] & divergence & --\\
40 & 8.80 879 102 228 926 [$-$32] & divergence & --\\
50 & 5.68 572 697 434 438 [$-$32] & divergence & --\\
60 & 3.96 925 492 004 272 [$-$32] & divergence & --\\
\hline
\multicolumn{4}{c}{$\alpha_1=50.0$, $\alpha_2=50.0$, $p=0$} \\
\hline
30 & 3.80 314 546 909 033 [$-$49] & divergence & --\\
40 & 2.20 243 367 882 466 [$-$49] & divergence & --\\
50 & 1.43 106 434 831 316 [$-$49] & divergence & --\\
60 & 1.00 278 875 314 209 [$-$49] & divergence & --\\
\end{longtable}
\begin{longtable}{cccc}
\caption[aa]{The functions $W_\mu(p;\alpha_1,\alpha_2)$ calculated for some representative
values of the parameters. \emph{Exact} denotes values calculated using explicit expressions (\emph{c.f.} Ref.
[\onlinecite{lesiuk2014b}]) in
extended arithmetic precision with the \textsc{Mathematica} package (all digits shown are correct). \emph{Large-order
expansion} column shows results of calculations with Eq. (\ref{lmuasym}) in the double precision arithmetic.
\emph{Convergence} denotes a number of terms in Eq. (\ref{lmuasym}) required to converge the summation to relative
precision of $2\cdot10^{-16}$. The symbol [k] denotes the powers of 10, 10$^k$. } \label{table3} \\
\hline\hline
$\mu$ & exact & large-order expansion & convergence \\
\hline \endfirsthead
\multicolumn{4}{l}%
{{\tablename\ \thetable{}: Continued from the previous page.}} \\
\endhead
\multicolumn{4}{r}{{Continued on the next page}} \\
\endfoot
\hline \hline
\endlastfoot
\multicolumn{4}{c}{$\alpha_1=1.0$, $\alpha_2=1.0$, $p=8$} \\
\hline
30 & 8.57 988 797 367 550 [$-$02] & 8.57 988 797 367 554 [$-$02] & 13\\
40 & 4.83 735 222 300 728 [$-$02] & 4.83 735 222 300 730 [$-$02] & 11\\
50 & 3.10 265 767 847 292 [$-$02] & 3.10 265 767 847 291 [$-$02] & 9\\
60 & 2.15 848 287 860 469 [$-$02] & 2.15 848 287 860 470 [$-$02] & 9\\
\hline
\multicolumn{4}{c}{$\alpha_1=1.0$, $\alpha_2=5.0$, $p=8$} \\
\hline
30 & 3.76 412 026 039 309 [$-$06] & 3.76 412 026 039 311 [$-$06] & 19\\
40 & 2.10 468 765 565 305 [$-$06] & 2.10 468 765 565 306 [$-$06] & 17\\
50 & 1.34 482 396 704 125 [$-$06] & 1.34 482 396 704 126 [$-$06] & 15\\
60 & 9.33 659 812 480 290 [$-$07] & 9.33 659 812 480 295 [$-$07] & 13\\
\hline
\multicolumn{4}{c}{$\alpha_1=1.0$, $\alpha_2=10.0$, $p=8$} \\
\hline
30 & 4.37 682 376 980 937 [$-$09] & 4.37 682 376 980 939 [$-$09] & 29\\
40 & 2.45 407 770 819 367 [$-$09] & 2.45 407 770 819 369 [$-$09] & 21\\
50 & 1.57 011 062 044 134 [$-$09] & 1.57 011 062 044 135 [$-$09] & 19\\
60 & 1.09 084 224 914 003 [$-$09] & 1.09 084 224 914 004 [$-$09] & 19\\
\hline
\multicolumn{4}{c}{$\alpha_1=1.0$, $\alpha_2=50.0$, $p=8$} \\
\hline
30 & 1.79 779 640 123 314 [$-$27] & divergence & $-$ \\
40 & 1.01 180 529 453 933 [$-$27] & divergence & $-$ \\
50 & 6.48 466 217 722 202 [$-$28] & divergence & $-$\\
60 & 4.50 946 436 202 621 [$-$28] & 4.50 946 436 202 621 [$-$28] & 181\\
\hline
\multicolumn{4}{c}{$\alpha_1=5.0$, $\alpha_2=5.0$, $p=8$} \\
\hline
30 & 1.44 602 672 357 563 [$-$08] & 1.44 602 672 357 562 [$-$08] & 21 \\
40 & 8.19 252 305 004 658 [$-$09] & 8.19 252 305 004 661 [$-$09] & 17 \\
50 & 5.26 663 146 433 895 [$-$09] & 5.26 663 146 433 898 [$-$09] & 15\\
60 & 3.66 849 871 467 375 [$-$09] & 3.66 849 871 467 377 [$-$09] & 13\\
\hline
\multicolumn{4}{c}{$\alpha_1=5.0$, $\alpha_2=10.0$, $p=8$} \\
\hline
30 & 4.24 176 585 818 655 [$-$11] & 4.24 176 585 818 656 [$-$11] & 29 \\
40 & 2.40 082 417 196 577 [$-$11] & 2.40 082 417 196 578 [$-$11] & 21 \\
50 & 1.54 267 805 985 739 [$-$11] & 1.54 267 805 985 739 [$-$11] & 19\\
60 & 1.07 428 909 373 431 [$-$11] & 1.07 428 909 373 431 [$-$11] & 19\\
\hline
\multicolumn{4}{c}{$\alpha_1=5.0$, $\alpha_2=50.0$, $p=8$} \\
\hline
30 & 2.98 140 021 972 450 [$-$29] & divergence & $-$ \\
40 & 1.68 614 952 443 023 [$-$29] & divergence & $-$ \\
50 & 1.08 306 875 847 441 [$-$29] & divergence & $-$ \\
60 & 7.54 082 251 283 390 [$-$30] & divergence & $-$ \\
\hline
\multicolumn{4}{c}{$\alpha_1=10.0$, $\alpha_2=10.0$, $p=8$} \\
\hline
30 & 1.75 800 825 092 494 [$-$13] & 1.75 800 825 092 495 [$-$13] & 29 \\
40 & 9.99 225 776 925 089 [$-$14] & 9.99 225 776 925 094 [$-$14] & 21 \\
50 & 6.43 335 525 002 640 [$-$14] & 6.43 335 525 002 643 [$-$14] & 19\\
60 & 4.48 491 212 960 371 [$-$14] & 4.48 491 212 960 374 [$-$14] & 19\\
\hline
\multicolumn{4}{c}{$\alpha_1=10.0$, $\alpha_2=50.0$, $p=8$} \\
\hline
30 & 1.79 687 571 912 046 [$-$31] & divergence & $-$ \\
40 & 1.02 110 421 809 320 [$-$31] & divergence & $-$ \\
50 & 6.57 355 034 007 000 [$-$32] & divergence & $-$ \\
60 & 4.58 239 720 894 004 [$-$32] & divergence & $-$ \\
\hline
\multicolumn{4}{c}{$\alpha_1=50.0$, $\alpha_2=50.0$, $p=8$} \\
\hline
30 & 4.14 652 002 770 361 [$-$49] & divergence & $-$ \\
40 & 2.39 745 614 502 924 [$-$49] & divergence & $-$ \\
50 & 1.55 652 302 599 839 [$-$49] & divergence & $-$ \\
60 & 1.09 019 745 763 500 [$-$49] & divergence & $-$ \\
\end{longtable}
\end{center}

\end{document}